\documentclass[pra,twocolumn,showpacs,preprintnumbers]{revtex4}

\usepackage{ifthen}
\usepackage{ifpdf}
\usepackage{color}

\ifpdf
\usepackage{graphicx}
\usepackage{epstopdf}
\else
\usepackage{graphicx}
\usepackage{epsfig}
\fi

\usepackage{latexsym}
\usepackage{amsmath}
\usepackage{amssymb}
\usepackage{bm}
\usepackage{wasysym}

\newcommand{\ei}{\hat{a}}
\newcommand{\eidag}{\hat{a}^{\dag}}

\newcommand{\Jx}{\hat{J}_x}
\newcommand{\Jy}{\hat{J}_y}
\newcommand{\Jz}{\hat{J}_z}

\newcommand{\Ji}{\hat{J}_i}
\newcommand{\Jj}{\hat{J}_j}

\newcommand{\sx}{R_x}
\newcommand{\sy}{R_y}
\newcommand{\sz}{R_z}
\newcommand{\sv}{{\bf R}}

\newcommand{\Dxx}{\Delta_{xx}}
\newcommand{\Dyy}{\Delta_{yy}}
\newcommand{\Dzz}{\Delta_{zz}}
\newcommand{\Dxy}{\Delta_{xy}}
\newcommand{\Dyz}{\Delta_{yz}}
\newcommand{\Dxz}{\Delta_{xz}}

\newcommand{\beq}{\begin{equation}}
\newcommand{\eeq}{\end{equation}}
\newcommand{\beqa}{\begin{eqnarray}}
\newcommand{\eeqa}{\end{eqnarray}}

\newcommand{\ra}{\rangle}
\newcommand{\cn}{{\rm cn}}

\newcommand{\hide}[1]{\textcolor{red}{[hidden]}}

\begin{document}

\title{Robust sub-shot-noise measurement via Rabi-Josephson oscillations in bimodal Bose-Einstein condensates}

\author{I. Tikhonenkov$^1$, M. G. Moore$^2$,  and A. Vardi$^{1}$}
\affiliation{$^1$Department of Chemistry, Ben-Gurion University of the Negev, P.O.B. 653, Beer-Sheva 84105, Israel\\
$^2$Department of Physics \& Astronomy, Michigan State Univerity, East Lansing, Michigan 48824, USA
}

\begin{abstract}
Mach-Zehnder atom interferometry requires hold-time phase-squeezing to attain readout accuracy below the standard quantum limit.  This increases its sensitivity to phase-diffusion, restoring shot-noise scaling of the optimal signal-to-noise ratio, $s_o$, in the presence of interactions. The contradiction between the preparations required for readout accuracy and robustness to interactions,  is removed by monitoring Rabi-Josephson oscillations instead of relative-phase oscillations during signal acquisition. Optimizing $s_o$  with a Gaussian squeezed input, we find that hold-time number squeezing satisfies both demands and that sub-shot-noise scaling is retained even for strong interactions. \end{abstract}


\maketitle

\section{Introduction}
Bimodal Bose-Einstein condensates (BECs), realized via double-well confinement \cite{Shin04,Albiez05,Schumm05,Est08} or internal spin states \cite{Boehi09,Gross10,Riedel10},  hold great promise for high-precision measurements. Their long phase coherence times and controllability of coupling and interaction parameters, make them ideal for the construction of atom interferometers which will make use of quantum correlations to reach unprecedented accuracy.

The current paradigm for matter-wave interferometers is the separated-pulses Mach-Zhender (MZ) scheme. The bimodal input state is mixed by a 50/50 beam-splitter, then held for a fixed duration of relative-phase acquisition in the presence of inter-mode bias, and then mixed again by a second 50/50 beamsplitter. The output number-difference then reflects the accumulated phase differential and through it, the bias present during the hold-time. 

For a two-mode coherent input, the interferometer phase uncertainty, $\Delta\theta$ is limited
by the Standard Quantum Limit (SQL),  $\Delta\theta=1/\sqrt{N}$, with $N$ being the total number of particles used. The SQL is also known as the 'shot-noise limit' because it is essentially the classical expectation for a 50/50 ensemble subject to $N$ measurements. By contrast, the preparation of strongly correlated, number-sqeezed input states, can push the phase-estimation uncertainty further down towards the Heisenberg limit $\Delta\theta= 1/N$, which is a factor $\sqrt{N}$ below shot-noise \cite{Bou97,Pezze}, as demonstrated in two recent experiments \cite{Gross10,Riedel10}.  In these experiments controlled interactions were used to dynamically generate squeezing via a one-axis twist strategy\cite{Kitagawa93}.

While strong interaction is essential for initial number-squeezing, it also limits the precision of the interferometer due to phase-diffusion during the phase acquisition time \cite{Jo07,Grond10}. To the lowest order, phase-diffusion is the shearing of the initial state due to the relative-number dependent mean-field shift. Hence  its rate is proportional to the number-variance \cite{Jo07} and the phase-squeezing required to attain sub-shot-noise accuracy, increases the sensitivity to interactions during the hold time. By contrast, states which are number-squeezed during the phase-acquisition period, are far more robust, but suffer from inherently large readout uncertainty. 

One approach to tackle this interplay between readout uncertainty and robustness against phase-diffusion, is to search for the optimal squeezing and phase-acquisition time which will yield the best accuracy for a given interaction strength and particle number \cite{Grond10,Huang08,Tikhonenkov10}. In previous work \cite{Huang08,Tikhonenkov10}, we have carried out such optimization for Gaussian Squeezed States (GSS), which constitute an excellent approximation to the ground-state of the bimodal BEC at T=0, with the squeezing controlled by adiabatic variation of the interaction-to-tunneling ratio \cite{Imamoglu97}. The GSS also closely approximates the states formed dynamically by the nonlinear beam-splitting of Refs.~\cite{Gross10,Riedel10}.  The non-interacting case was considered in \cite{Huang08} and the effect of interactions was later studied in \cite{Tikhonenkov10}. In the latter work \cite{Tikhonenkov10}, we adopted the view that the bimodal BEC MZ interferometer measures the `bias', or energy differential, $\varepsilon$, between the two wells, rather than the accumulated phase-shift $\theta=\varepsilon T$. Thus the hold time $T$ was taken as a free parameter, which can be used in supplement to the initial number-difference variance $\sigma$ to optimize the signal-to-noise ratio (SNR) $s=\Delta\varepsilon/\varepsilon$ of the interferometer, at any fixed value of $N$ and the dimensionless interaction parameter $\eta=UN/(2\varepsilon)$, where $U$ is the interaction strength.

Using this approach, we found that the optimal preparation changes from a number-squeezed state (transformed to hold-time phase-squeezed state by the beam-splitter) with $\sigma_o\sim N^{1/3}$ in the absence of interactions \cite{Huang08} to a phase-squeezed state with $\sigma_o\sim \sqrt{\eta N}$ (here and below `o' subscripts indicate optimal values) for strong interactions \cite{Tikhonenkov10}. The resulting optimal SNR is subject to a similar transition from the better than shotnoise scaling $s_o\sim N^{2/3}$ without interactions to the strong-interaction behavior of $s_o\sim \sqrt{N/\eta}$. Thus sub-shotnoise precision is lost in the MZ scheme due to interactions and the optimal SNR is worse by a factor $\sqrt{\eta}$ than the standard quantum limit. 

\begin{figure}[t]
\centering
\includegraphics[angle=90,width=0.45\textwidth]{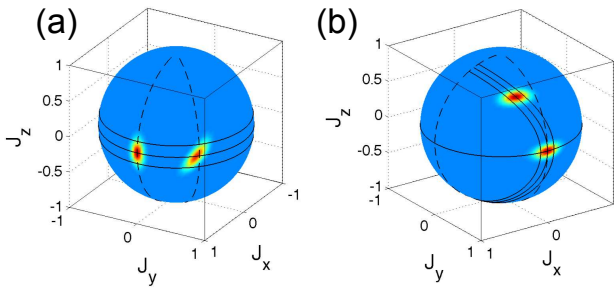}
\caption{(Color online) Signal acquisition stage for (a) Mach-Zehnder interferometer and (b) the proposed Rabi-Josephson scheme. Husimi distributions are plotted immediately after the 'beam-splitting' and at the end of the hold time. Solid lines depict the classical, mean-field trajectories.}
\label{Fig1_schematic}
\end{figure}

Here we consider a different strategy in order to altogether avoid the contradiction between the requirements for read out accuracy (namely hold-time phase squeezing)  and robustness to phase-diffusion (i.e. hold-time number-squeezing). To achieve this goal we replace the roles of the bias $\varepsilon$ and hopping $K$ during the signal acquisition stage of the interferometer. Thus, instead of using phase-oscillations to monitor $\varepsilon$ we propose to directly measure the nonlinear Rabi-Josephson amplitude oscillation in order to determine $K$ (see Fig.~\ref{Fig1_schematic}).  This way, we expect that the preferred preparation for readout accuracy involves hold-time {\it number}-squeezing instead of phase-squeezing. Consequently the optimal input states should not lead to faster phase-diffusion but in fact be more robust against interactions. Repeating the optimization process, we find  that the optimal squeezing scales as $\sigma_o\sim(\tau_o N)^{1/3}$ where $\tau_o=KT_o\sim u^{-1/2}$ and $u=UN/(2K)$ is the pertinent interaction parameter. The optimal acquisition angle $\tau_o$ is only residually dependent on $N$. Consequently, the optimal SNR $s=K/\Delta K$ scales as $s_o\sim (\tau_o N)^{2/3}$. Therefore, stronger phase-squeezing $\sigma_o\sim u^{-1/6}$ is preferred for stronger interactions and sub-shotnoise scaling $s_o\sim u^{-1/3} N^{2/3}$ is maintained.

Experimentally, current double-well setups rely on the combination of a 3D harmonic trap and a 1D optical lattice,
\begin{equation}
V(x,y,z)=\sum_{q=x,y,z} \omega_q q^2+V_0\cos^2(k_0 z),
\end{equation}
where $\omega_{x,y,z}$ are the trap frequencies along the respective axes and $k_0$ is the optical lattice wavenumber. Mach-Zendher interferometry is implemented to measure the tilt $V_T(z)=E_z z$ introduced by a field  applied along the lattice axis. If instead a field $E_x$ is applied perpendicular to the axis, its effect would be shift the harmonic trap-center along $x$. Provided that $V_0$ is made $x$-dependent (e.g. by using a 2D lattice whose wavelength along $x$ is long compared to the harmonic trap size or by implementing a perpendicular $V_0$ gradient so that $V_0\to V_0(x)$), the barrier height and width would be sensitive to the magnitude of the perpendicular field $E_x$ and could be used to measure it. Such a device would thus be sensitive to parallel field gradients through variations of $\varepsilon$ measured via relative-phase oscillations and to perpendicular field gradients through variations of $K$ measured via relative-amplitude Rabi-Josephson oscillations.

\section{Measurement Scheme and Optimization procedure}
As in previous work, we use the bimodal approximation of the two-site Bose-Hubbard Hamiltonian \cite{Vardi01}, which accounts for the pertinent physics in current experimental setups \cite{Li09},
\begin{equation}
\label{Ham}
H=-K\Jx+{\varepsilon}\Jz+U\Jz^2~.
\end{equation}
Here $K$, $\varepsilon$, and $U$ are coupling, bias, and interaction energies, where $U>0$ corresponds to repulsive interactions and vice versa.  The bias $\varepsilon$ may be positive or negative, depending on the energy detuning between the two modes. The SU(2) generators $\Jx=(\eidag_1 \ei_2+\eidag_2\ei_1)/2$,  $\Jy=(\eidag_1\ei_2-\eidag_2\ei_1)/(2i)$, and $\Jz=(n_1 - n_2)/2$, are defined in terms of the boson on-site annihilation and creation operators $\ei_i$, $\eidag_i$, with the conserved total particle number $n_1+n_2=N\equiv 2j$.

The standard MZ scheme consists of a fast $\pi/2$ beam-splitter rotation about $J_x$, followed by relative-phase acquisition during a hold-time $T$ due to the bias detuning $\varepsilon$, and an opposite $\pi/2$ readout rotation about $J_x$. The final population imbalance $J_z^f$  is used to read the accumulated phase $\theta=\varepsilon T$ from which the bias, $\varepsilon$, is readily obtained.  Assuming that the beam-splitter and read-out rotations are instantaneous with respect to the characteristic phase-diffusion time, the MZ interferometer can be described by the propagator
\beq
U_{MZI}(\theta, \eta,j)=e^{-i\frac{\pi}{2} \Jx }e^{-i\theta\Jz\left(1-(\eta/j)\Jz\right)} e^{-i\frac{\pi}{2}\Jx},
\label{UMZI}
\eeq
where $\eta=Uj/\varepsilon$. The simultaneous operation of phase-diffusion and phase acquisition during the hold time $T$ degrades the accuracy of the interferometer, as described in \cite{Tikhonenkov10,Grond10}.

Instead, we propose to employ a Rabi-Josephson (RJ) scheme, summed by the propagator
\beq
U_{RJ}(\tau, u,j)=e^{-i\tau\left(\Jx-\frac{u}{j}\Jz^2\right)} e^{-i\frac{\pi}{2}\Jz}
\label{URJ}
\eeq
where $u=Uj/K$ and $\tau=KT$, in order to determine the value of the coupling strength $K$ from the final population difference. As in our analysis of the MZ interferometer \cite{Huang08,Tikhonenkov10}, this propagator acts on a Gaussian squeezed state, of the form,
\begin{equation}
|\sigma\rangle=\frac{1}{\sqrt{{\cal N}_\sigma}}\sum_{m=-j}^j \exp\left[-\frac{m^2}{4\sigma^2}\right]|j,m\rangle~,
\label{sigma}
\end{equation}
where $\sigma$ is the initial number-difference uncertainty, and ${\cal N}_\sigma=\sum_{m=-j}^j \exp(-m^2/(2\sigma^2))\approx\sqrt{2\pi}\sigma$.  Such states approximate well the ground state of  Hamiltonian (\ref{Ham}) with $U>0$ and $\varepsilon\approx0$ \cite{Imamoglu97}, as well as the  dynamically squeezed preparations observed in recent experiments \cite{Kitagawa93,Gross10,Riedel10}.

In contrast to the MZ interferometer, where the commutation of the operators for phase-acquisition and phase-diffusion allows for separation of the two processes and the derivation of analytic expressions, we now have to carry out the simultaneous evolution under $\Jx$ and $\Jz^2$. This amounts to the nonlinear mean-field Bose-Josephson oscillation, accompanied by the deformation of the initial distribution due to the variation of initial conditions.   We evaluate the final population difference $J_z^f=\langle \Jz\rangle_{T}$ and its variance $(\Delta J_z^f)^2=\langle \Jz^2\rangle_T-\langle \Jz\rangle_T^2$. Using the error propagation estimate $\Delta K = \Delta J_z^f/(\partial J_z^f /\partial K)$, we find the the SNR ratio is given by, 
\begin{equation}
s=\frac{K}{\Delta K}=\left | \frac{\partial J_z^f}{\partial\tau}\right|\frac{\tau}{\Delta J_z^f}~.
\label{SNR}
\end{equation}

This quantity is optimized with respect to the parameters, $\{\sigma,\tau\}$, in order to obtain the maximum signal-to-noise ratio. Because the final state $|\sigma,\tau\ra\equiv U_{RJ}|\sigma\ra$ depends only on the parameters $\{\tau,u,j,\sigma\}$, it follows that the optimal values, $(\sigma_o,\tau_o)$ that give the maximum SNR,  $s_o=s(\sigma_o,\tau_o)$, as well as $s_o$ itself, are functions of the particle number, $N=2j$ and the interaction-to-tunneling ratio $u$, only. Our goal is to determine the scaling of $s_o$ with $j$ and $u$ and to establish if sub-shotnoise scaling of $s_o$ is maintained in the presence of interactions.

\begin{figure}[t]
\centering

\includegraphics[angle=0,width=0.5\textwidth]{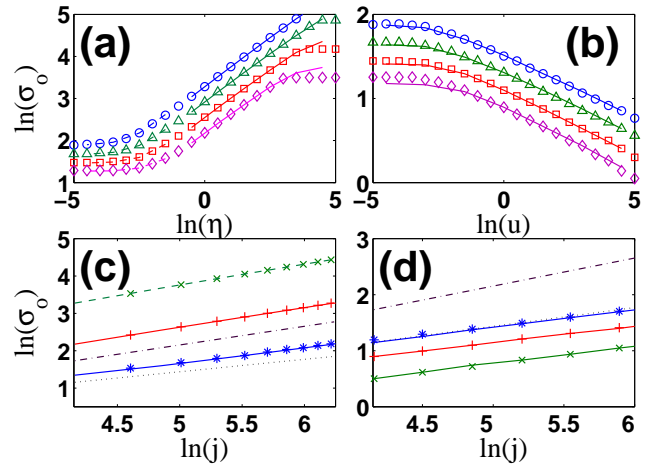}
\caption{(Color online) Optimal initial number variance $\sigma_o$ required to maximize the SNR: (a) as a function of interaction-strength $\eta$ at fixed $j=64$, ($\diamond$), $128$ ($\square$), $256$ ($\triangle$), and 512 ($\circ$) for a MZ interferometer; (b) same for the proposed RJ scheme, with $u$ being the relevant hold-time interaction parameter; (c) as a function of the atom number $j$ at fixed interaction strength $\ln(\eta)=-2.5$ ($*$), 0 ($+$), and $+2.5$ ($\times$) for the MZ interferometer; (d) same for RJ scheme with same values of $\ln(u)$ using the same notation convention. Symbols in all panels denote full numerical simulation results. Lines in (a,c) correspond to analytic expressions obtained in Ref.~\cite{Tikhonenkov10}  for weak interaction (dashed) and strong interaction (solid).  Solid lines in (b,d) correspond to the BBR truncated-cumulant-expansion \cite{Tikhonenkov07}. Dotted lines in panels (c),(d) mark the interaction-free optimal-squeezing  scaling $\sigma_o=j^{1/3}$ whereas dash-dotted lines mark the coherent-state (classical) variance $\sigma=\sqrt{j}$} 
\label{Fig2_sigmaopt}
\end{figure}


\begin{figure}[t]
\centering
\includegraphics[angle=0,width=0.5\textwidth]{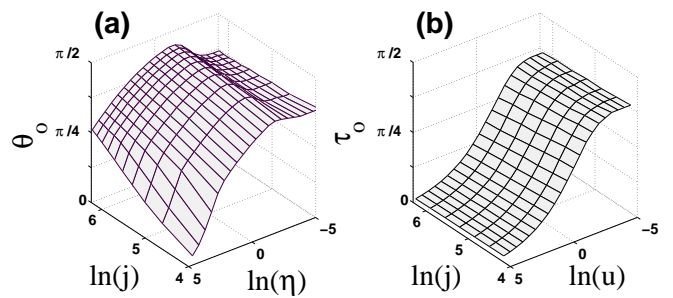}
\caption{Optimal acquired angle $\theta_o=\varepsilon T_o$, $\tau_o=KT_o$ required to maximize the SNR for (a) the MZ and (b) the RJ schemes, respectively, as a function of the pertinent interaction parameter $\eta,u$ and the particle number $j$. } 
\label{Fig3_tauopt}
\end{figure}

\section{Numerical Optimization Results}

\begin{figure}[t]
\centering
\includegraphics[angle=0,width=0.5\textwidth]{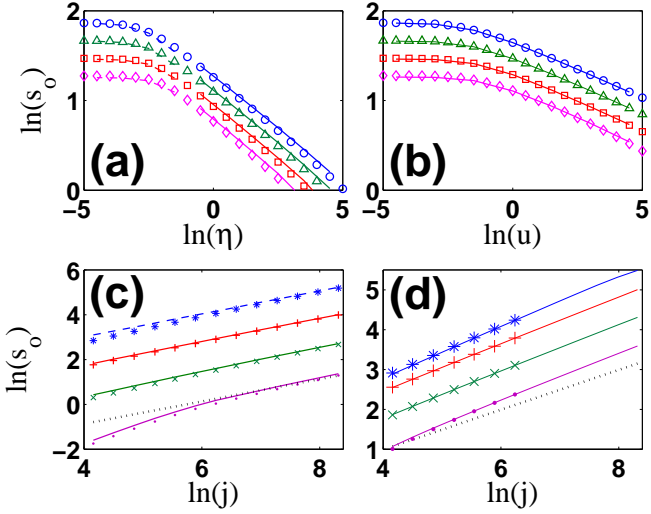}
\caption{(Color online) Optimal precision $p_o\equiv\ln s_o$. Panels are arranged in the same order as in Fig.~\ref{Fig2_sigmaopt}. Values of $j$ in Panels (a),(b) are also the same.  In (c) and (d), the values of $\ln(\eta)$ and $\ln(u)$, respectively, are  -2.5 ($*$), 0 ($+$), 2.5 ($\times$), and 5 ($\cdot$). Using the same notation as in Fig.~\ref{Fig2_sigmaopt}, markers denote numerics, whereas lines are analytic estimates (a,c) and BBR approximate values (b,d). Dotted lines in (c),(d) mark shot-noise scaling $s\sim j^{1/2}$.}
\label{Fig4_SNRopt}
\end{figure}

In Fig.~\ref{Fig2_sigmaopt} we plot the results of a numerical evaluation (solid lines) of the optimal squeezing $\sigma_o$ required to obtain the best SNR in both the MZ (Fig.~\ref{Fig2_sigmaopt}a,c) and RJ (Fig.~\ref{Fig2_sigmaopt}b,d) schemes, as a function of the pertinent interaction parameters ($\eta$ for MZ or $u$ for RJ) and particle number $j$. Symbols correspond to the analysis oulined below in Section IV. For both MZ and RJ schemes, the optimal squeezing, as well as the resulting SNR (Fig.~\ref{Fig4_SNRopt}), coincide in the weak interaction regime $\eta,u<1/2$, with the previously studied interaction-free MZ propagation \cite{Huang08}. This is expected because without interaction the choice of axes is immaterial and the  RJ sequence may be viewed as a simple rotation of the standard MZ interferometer. For stronger $\eta,u>1/2$ interactions (corresponding to the Josephson regime \cite{Smerzi97} for the RJ scheme) however, significant differences are observed.

For the MZ sequence, the optimal initial number variance increases with increasing interactions, until a transition from optimal initial number-squeezing ($\sigma_o<\sqrt{j/2}$) to optimal initial phase-squeezing ($\sigma_o>\sqrt{j/2}$) takes place as $\eta$ crosses unity. This transition results from the said contradiction between projection noise minimization (initial number squeezing, hold-time phase-squeezing) and phase-diffusion control (initial phase-squeezing, hold-time number-squeezing). Thus, while for weak interaction ($\eta\ll 1$), we obtain sub-Poissonian scaling $\sigma_o\propto j^{1/3}$ \cite{Huang08},  in the presence of strong interactions ($\eta> 1$)  we have super-Poissonian optimal-input number distribution with $\sigma_o\propto\sqrt{\eta N}$ \cite{Tikhonenkov10}. In comparison, the optimal squeezing in the RJ scheme does not follow a similar compromise. Initial number squeezing is preferred for both weak and strong interactions, and in fact becomes stronger with increasing the value of the interaction strength $u$. The optimal squeezing in this case, is found numerically to scale as $(\tau_o N)^{1/3}$.

Significant differences between the standard and proposed schemes also appear in the behavior of the optimal acquired angle ($\theta_o=\varepsilon t_o$ for MZ  and $\tau_o=Kt_o$ for RJ), shown in Fig.~\ref{Fig3_tauopt}. For negligible interactions, both are optimized by $\sim 3\pi/8$ \cite{Tikhonenkov10} as the pure Rabi oscillation is a simple rotation of the phase-oscillations of the Mach-Zehnder interferometer.  However, the $j$ dependence of the optimal $\tau$ in the RJ scheme in the presence of interactions is much weaker than that of $\theta_o$ for the MZ interferometer. The resulting maximized precision $p_o=\ln_{10}s_o$ ($p_o$ corresponds to the number of significant read-out figures) is plotted in Fig.~\ref{Fig4_SNRopt}, using the same conventions and ordering. For the MZ scheme, the weak-interaction sub-shot-noise scaling $s_o\propto j^{2/3}$ \cite{Huang08} is replaced by $s_o\propto\sqrt{N/\eta}$ for $\eta>1$, i.e. worse than the standard quantum limit. The RJ scheme is far more robust to interaction and the SNR deterioration is slower than $s_o\propto u^{-1/3}$ as compared to $s_o\propto u^{-1/2}$ for the MZ scheme \cite{Tikhonenkov10}. Consequently, the RJ scheme  {\it retains sub-shot-noise scaling}, with $s_o\sim(\tau_o N)^{2/3}$. 

\section{Bogoliubov Analysis}
For the MZ scheme we were able to derive exact analytic expressions for the dynamics with general Gaussian initial conditions and extract the pertinent scaling laws for $\sigma_o$ and $s_o$ by retaining leading order terms in the various interaction regimes \cite{Tikhonenkov10,Grond10}.  Such derivation is not possible for the proposed RJ scheme  due to the incommutability of the $\Jx$ and $\Jz^2$ terms during the acquisition stage. In order to analytically prove the numerically-observed scaling relations of the RJ scheme, we employ a density-matrix cumulant-expansion technique developed under the name Bogoliubov Back-Reaction (BBR) \cite{Tikhonenkov07}.  The hierarchy of dynamical equations for the $\Ji$ operators, is truncated at second-order to obtain equations of motion for the mean-field single-particle Bloch vector $\sv\equiv\langle\hat{\bf J}\rangle/j$ and the correlation functions $\Delta_{ij}=(\langle\Ji\Jj+\Jj\Ji\rangle-2\langle\Ji\rangle\langle\Jj\rangle)/j^2$. Neglecting the back-action terms, these read,
\begin{eqnarray}
\dot\sx&=&-2u \sy \sz,\nonumber\\
\dot\sy&=&s_z+2u \sx \sz,\nonumber\\
\dot\sz&=&-\sy\nonumber\\
{\dot \Delta}_{xx}&=&-4u\left(\sy\Delta_{xz}+\sz\Delta_{xy}\right),\nonumber\\
{\dot \Delta}_{yy}&=&2\Delta_{yz}+4u\left(\sx\Delta_{yz}+\sz\Delta_{xy}\right),\nonumber\\
{\dot \Delta}_{zz}&=&-2\Delta_{yz}\nonumber\\
{\dot \Delta}_{xy}&=&\Delta_{xz}+2u \sz\left(\Delta_{xx}-\Delta_{yy}\right)+2u\left(\sx\Delta_{xz}-\sy\Delta_{yz}\right),\nonumber\\
{\dot \Delta}_{xz}&=&-\Delta_{xy}+2u \left(\sy\Delta_{zz}+\sz\Delta_{yz}\right),\nonumber\\
\label{BBR}
{\dot \Delta}_{yz}&=&\Delta_{zz}-\Delta_{yy}+2u \left(\sx\Delta_{zz}+\sz\Delta_{xz}\right),
\end{eqnarray}
where $\dot{f}\equiv d{f}/d\tau$. 

Given the values of $\sv$ and $\Delta_{ij}$ after the preliminary $\pi/2$ rotation about $J_z$,
\begin{equation}
\sv_0=(0,1,0)~,~\Delta_{xy,0}=\Delta_{xz,0}=\Delta_{yz,0}=0,\nonumber
\end{equation}
\begin{equation}
\Delta_{xx,0}\approx\frac{1}{2\sigma^2}~,~\Delta_{yy,0}\approx\frac{1}{16\sigma^4}~,~\Delta_{zz,0}\approx\frac{2\sigma^2}{j^2},
\label{initial}
\end{equation}
we obtain an exact solution to Eqs.~(\ref{BBR}) in terms of elliptical Jacobi functions (see Appendix). To leading order, we find that,
\begin{eqnarray}
\left(\Delta J_z^f\right)^2/j^2=\Delta_{zz}(\tau)/2&\approx&\frac{u^2\tau^6}{36\sigma^2}+\frac{\tau^2}{32\sigma^4}+\frac{\sigma^2}{j^2},\\
\frac{1}{j}\left|\frac{\partial J_z^f}{\partial\tau}\right|=\left| R_y(\tau)\right|&\approx&\sqrt{1-\tau^2-u^2\tau^4}.
\end{eqnarray}
Substituting into Eq.~(\ref{SNR}) and focusing on the strong interaction regime $u\tau\gg1$, we obtain,
\begin{equation}
s^2(\tau,\sigma)=\frac{\tau^2\left(1-u^2\tau^4\right)}{\frac{u^2\tau^6}{36\sigma^2}+\frac{\tau^2}{32\sigma^4}+\frac{\sigma^2}{j^2}}.
\label{SNRBBR}
\end{equation}
Optimizing $s(\tau,\sigma)$ with respect to $\tau$ and $\sigma$, results in
\begin{equation}
u^2\tau^4\sigma^2=A(u,\tau)~,~\sigma^2=B(u,\tau)j^{2/3}\tau^{2/3},
\end{equation}
where the functions
$$
A(u,\tau)=\frac{9}{4}\frac{1-4u^2\tau^4}{1+3u^2\tau^4}~,~B(u,\tau)=\left( \frac{1}{16}+\frac{A(u,\tau)}{36}\right)^{1/3},
$$
only depend weakly on $u$ and $\tau$ and could therefore be assumed constant. Thus we conclude that the scaling of the optimal initial number variance $\sigma_o$ and hold time $\tau_o$ required to obtain the best SNR $s$ in the RJ scheme, is
\begin{equation}
\sigma_o^2=(j\tau_o)^{2/3}/2~,~u^2\tau_o^4\sigma_o^2=9/4~.
\label{best}
\end{equation}
This is consistent with the numerically obtained $\sigma_o\sim u^{-1/6} j^{1/3}$, $s_o\sim j^{2/3}u^{-1/3}$ scaling in Figs.~\ref{Fig2_sigmaopt},\ref{Fig4_SNRopt}. The results of the BBR optimization are shown as symbols in Figs.~\ref{Fig2_sigmaopt}(b),(d) and Figs.~\ref{Fig4_SNRopt}(b),(d) and agree well with the numerics. 

\section{Conclusions}
Due to their strong interaction, phase-diffusion is an eminent obstacle to the realization of sub-shotnoise interferometry in dilute quantum gases. The standard Mach-Zehnder scheme borrowed from linear optical interferometry, suffers from an inherent contradiction between the preparation of a number-squeezed input state required for sub-shotnoise precision, and its increased sensitivity to interactions during the signal acquisition time after it has been converted to a phase-squeezed state by the beam-splitter \cite{Tikhonenkov10,Grond10}. Placing the measured perturbation between the sites so it affects the odd-even detuning instead of the bias between sites, and monitoring Rabi-Josephson amplitude oscillations instead of phase-oscillations, removes this contradiction. We have shown that number-squeezed states satisfy both readout and robustness requirements and that in fact increased number squeezing optimizes the input states in the presence of interactions. This is distinct from the MZ scheme, where a transition occurs from optimal hold-time phase-squeezing (optimizing readout) to optimal number-squeezing (optimizing robustness) as the interactions cross a critical magnitude \cite{Tikhonenkov10}. As a result, the best SNR ratio obtained by the RJ scheme retains sub-shotnoise scaling, as opposed to the super-shotnoise scaling of the SNR of a MZ interacting atom interferometer.

\appendix
\section{Solution of the BBR equations}
Given the set of equations (\ref{BBR}) with the initial conditions (\ref{initial}), an exact analytic solution for the mean-field equations for $\sv$ is found in the form of Jacobian elliptic functions,
\begin{eqnarray}
\sx&=&u\sz^2,\nonumber\\
\sy&=&\sqrt{1-\sz^2-u^2\sz^4},\nonumber\\
\sz&=&-R_m\cn\left(\tau\left(1+4u^2\right)^{1/4}-K(k)\right),
\end{eqnarray}
where $R_m^2=2/(1+\sqrt{1+4u^2})$, $k^2=\left(1-1/\sqrt{1+4u^2}\right)/2$, $K(k)=\int_{0}^{\pi/2}d\phi\left(1-k^2\sin^2\phi\right)^{-1/2} $ is the quarter period, and
$$
\tau=\int_0^{|\sz|} \frac{d\zeta}{\sqrt{1-\zeta^2-u^2\zeta^4}}~.
$$

The equations for the fluctuations $\Delta_{ij}$, may then be solved in the form,
\begin{equation}
\left(
\begin{array}{ccc}
\Dxx&\Dxy&\Dxz\\
\Dxy&\Dyy&\Dyz\\
\Dxz&\Dyz&\Dzz
\end{array}
\right)=
C^T
\left(
\begin{array}{ccc}
\Delta_{xx,0}&0&0\\
0&\Delta_{yy,0}&0\\
0&0&\Delta_{zz,0}
\end{array}
\right)
C
\end{equation}
where the elements $c_{ij}$ of the orthogonal matrix $C$ satisfy,
\begin{eqnarray}
{\dot c}_{1a}&=&-2u\sz c_{2a}-2u\sy c_{3a},\nonumber\\
{\dot c}_{2a}&=&(1+2u\sx) c_{3a}+2u\sz c_{1a},\nonumber\\
{\dot c}_{3a}&=&-c_{2a}, 
\end{eqnarray}
with initial values $c_{ab}(t=0)=\delta_{ab}$ and $a,b=1,2,3$. Parametrizing the solutions in terms of the variable $\rho=-\sz$ and noting that,
\begin{equation}
\frac{dc_{ab}}{d\tau}=\frac{dc_{ab}}{d\rho}\sqrt{1-\rho^2-u^2\rho^4},
\end{equation}
we find the following series expansions:
\begin{eqnarray}
c_{11}&=&1-2u\sum_{n=1}^\infty\alpha_n\rho^{2n+2}~,\nonumber\\
c_{21}&=&-\sqrt{1-\rho^2-u^2\rho^4}\sum_{n=1}^\infty (2n+1)\alpha_n\rho^{2n+1},\nonumber\\
c_{31}&=&\sum_{n=1}^\infty\alpha_n\rho^{2n+1}~,\nonumber\\
c_{12}&=&2u\sum_{n=0}^\infty\frac{\beta_n}{2n+1}\rho^{2n+2}~,\nonumber\\
c_{22}&=&\sqrt{1-\rho^2-u^2\rho^4}\sum_{n=0}^\infty\beta_n\rho^{2n},\nonumber\\
c_{32}&=&-\sum_{n=0}^\infty\frac{\beta_n}{2n+1}\rho^{2n+1}~,\nonumber\\
c_{13}&=&-2u\sum_{n=0}^\infty\gamma_n\rho^{2n+1}~,\nonumber\\
c_{23}&=&-\sqrt{1-\rho^2-u^2\rho^4}\sum_{n=1}^\infty 2n\gamma_n\rho^{2n-1},\nonumber\\
c_{33}&=&\sum_{n=0}^\infty\gamma_n\rho^{2n}~.
\end{eqnarray}
The coefficients $\alpha_n,\beta_n,\gamma_n$ are given by the recursion relations,
$$\alpha_1=\frac{u}{3},~\alpha_2=\frac{2u}{15}~,$$
$$\alpha_{n+1}=\frac{2n}{2n+3}\alpha_n+u^2\frac{2n-3}{2n+3}\alpha_{n-1}
,~n\geq 2,$$
$$\beta_0=1,~\beta_1=0,~\beta_2=\frac{4u^2}{3},$$
$$\beta_{n+1}=\frac{2n(2n+2)}{(2n+1)^2}\beta_n+u^2\frac{(2n-3)(2n+2)}{(2n-1)(2n+1)}\beta_{n-1},~n\geq 2,$$
$$\gamma_0=1,~\gamma_1=-1/2$$
$$\gamma_{n+1}=\frac{2n-1}{2n+2}\gamma_n+u^2\frac{2n^2-3n+2}{(2n+1)(n+1)}\gamma_{n-1},~n\geq 1.$$


\begin{acknowledgements}
This work was supported  by the Israel Science Foundation (Grant 582/07), by grant no. 2008141 from the United States-Israel Binational Science Foundation (BSF), and by the National Science Foundation through a grant for the Institute for Theoretical Atomic, Molecular, and Optical Physics at Harvard University and Smithsonian Astrophysical Observatory.
\end{acknowledgements}

\end{document}